\documentstyle[12pt,aasms4]{article}
\def\etal   {{et~al.}\ }

\def\msun{{\rm\,M_\odot}}

\def\vol#1  {{{#1}{\rm,}\ }}
\def\lya{{\rm Ly}\alpha}

\def\etal{et al.\ }

\newcount\refno
\newcount\rfno
\def\eq{$^{\the\refno\ }$\advance\refno by 1}
\def\ad{\advance\rfno by 1}

\def\clock{\count0=\time \divide\count0 by 60
     \count1=\count0 \multiply\count1 by -60 \advance\count1 by \time
     \number\count0:\ifnum\count1<10{0\number\count1}\else\number\count1\fi}

\begin{document}
\title{Where are the Baryons?}
\author{Renyue Cen\altaffilmark{1} and Jeremiah P. Ostriker\altaffilmark{2}}
\altaffiltext{1} {Princeton University Observatory, Princeton University, Princeton, NJ 08544; cen@astro.princeton.edu}
\altaffiltext{2} {Princeton University Observatory, Princeton University, Princeton, NJ 08544; jpo@astro.princeton.edu}

\begin{abstract}

New, high resolution, large-scale, cosmological hydrodynamic 
galaxy formation simulations of a standard 
cold dark matter model (with a cosmological constant) are utilized
to predict the distribution of baryons at the present and at 
moderate redshift.
It is found that the average temperature of baryons is 
an increasing function of time,
with most of the baryons at the present time having 
a temperature in the range $10^{5-7}$K.
Thus, not only is the universe dominated by dark matter,
but more than one half of the normal matter is yet to be detected.
Detection of this warm/hot gas poses an observational challenge,
requiring sensitive EUV and X-ray satellites.
Signatures include a soft, cosmic X-ray background,
apparent warm components in hot clusters due to both
intrinsic warm intra-cluster gas and warm
inter-cluster gas projected
onto clusters along the line of sight,
absorption lines in 
X-ray and UV quasar spectra 
[e.g., O VI (1032,1038)A lines, OVII 574~eV line],
strong emission lines (e.g., O VIII 653~eV line)
and low redshift, broad, low column density $\lya$ absorption lines.
We estimate that approximately $1/4$ of the 
extragalactic soft X-ray background (SXRB) (at $0.7~$keV)
arises from the warm/hot gas,
half of it coming from $z<0.65$ and three-quarters from
$z<1.00$, so the source regions
should be identifiable on deep optical images.

\end{abstract}


\section{Introduction}

Where are the baryons, the ordinary non-exotic matter
of the universe?
If we consider the universe at redshift two,
a consistent picture emerges:
the amount of mass seen directly by the Hubble Space Telescope (HST)
in stellar systems is very small ($\Omega_*\approx 5.5\times 10^{-4}$
[for h=0.70]; Madau, Pozzetti, \& Dickinson 1998),
when compared either to the amount seen 
in the local universe or to the critical cosmological density.
But the amount of hydrogen and helium
observed as absorption lines in the Lyman alpha forest
is considerable.
Recent analyses (Rauch \etal 1997; Weinberg \etal 1997) give
\begin{equation}
\Omega_{baryon} \ge 0.017h^{-2} = 0.035
\end{equation}
\noindent 
from observations
of the Lyman alpha forest at redshift two,
where $\Omega_{baryon}$ is the density in units of the critical
density, $h\equiv H/100$km/s/Mpc,
and $h=0.70$ is adopted for the last term in equation (1).
It is worth noting that the above quoted lower limit on $\Omega_{baryon}$
might be too conservative by as much as 
a factor of $\sim 2-7$,  due primarily to 
the uncertainty in the observationally determined
metagalactic radiation field.

The observed light-element ratios,
combined with standard nucleosynthesis, allow to compute 
the expected baryon density for standard models (Burles \& Tytler 1998) as
\begin{equation}
\Omega_{baryon} = (0.019\pm 0.001)h^{-2} = 0.039\pm 0.002.
\end{equation}
\noindent 
The approximate consistency between these two 
completely independent methods of estimating the baryon density
is reassuring, as is the fact that
theoretical models incorporating (1), (2) and standard 
theory for the growth of structure 
reproduce, in great detail, numerous features
(column density dependence, redshift dependence, spatial correlations, etc)
of the observed Lyman alpha forests
(Cen \etal 1994; Zhang \etal 1995; Miralda-Escud\'e \etal 1995;
Hernquist \etal 1996).

But at redshift zero, in the present day universe,
every analysis (e.g., Fukugita, Hogan, \& Peebles 1997)
indicates that, after summing over all 
well observed contributions,
the local baryon density 
appears to be far lower than indicated by equation (1) and (2):
\begin{equation}
\Omega_{*} + \Omega_{HI} + \Omega_{H_2} + \Omega_{Xray,cl} \approx 0.0068 \le 0.011~~(2\sigma~\hbox{limit})
\end{equation}
\noindent 
for $h=0.70$.
Thus, either most of the baryons in the present day universe
are yet to be detected or a serious error
has been made in the arguments that led to equations (1) and (2).
Several authors (e.g., Wolfe \etal 1995) have
noticed that the mass density seen in the damped Lyman alpha
systems at moderate redshift ($z\sim 3$)
approximately corresponds to the mass density
in the stellar component of 
galaxies at $z=0$.
While important and roughly true,
this point does not
help us much,
since only a small fraction (approximately $4\%$)
of the Lyman alpha forest is seen in the 
neutral high column density damped systems and,
correspondingly, $\Omega_*$ is about $0.003$.

Thus, the puzzling question remains.
Where are the baryons in the universe today?
There have been a number of proposals to account for
the missing baryons (O'Neil 1997; Burkert \& Silk 1997; 
Bristow \& Phillips 1994).
Now, the whereabouts and state of those missing baryons
can be computed from standard initial 
conditions, in realistic large-scale
cosmological hydrodynamic simulations.
We have concluded
that 
a substantial fraction of the missing
baryons are to be found in intergalactic gas
in a temperature range $10^5<T<10^7$K, 
where they have been difficult to detect and 
which we call ``warm/hot gas".
The paper is organized as follows.
A brief descriptions of the simulations is presented in \S 2.
Results are given in \S 3
and \S 4 concludes the paper with a discussion of 
the robustness of the results and ways to detect 
this warm/hot gas.

\section{Simulation}

This result is based on a computation of 
the evolution of the gas in 
a cold dark matter model with a cosmological constant;
the model
is normalized to both the microwave background temperature
fluctuations measured by COBE 
on large scales (Bunn \& White 1997)
and the observed abundance of clusters of galaxies in
the local universe (e.g., Cen 1998),
and
it is close to both the concordance model of Ostriker \& Steinhardt (1995)
and the model indicated by the recent
high redshift supernova results (Riess \etal 1997).
The relevant model parameters
are: $\Omega_0=0.37$, $\Omega_{baryon}=0.049$,
$\Lambda_0=0.63$, $\sigma_8=0.80$,
$h=0.70$,
$n=0.95$ and $25\%$ tensor mode contribution
to the CMB fluctuations on large scales.
Two simulations with box sizes of $L_{box}=(100,50)h^{-1}$Mpc
are made, each having $512^3$ cells and $256^3$ dark matter particles.
The conclusions drawn in 
this paper are not significantly affected by the finite resolution.
Although based on simulations of a specific model,
we believe that the result is generic and would occur in any
model consistent with other large scale structure results.
The description of the
numerical methods of the cosmological hydrodynamic code 
and input physical ingredients will be presented elsewhere.
To briefly recapitulate,
we follow three components separately and simultaneously:
dark matter, gas and galaxies,
where the last component is created continuously from the former two
during the simulations in places where real galaxies are thought
to form as dictated mostly by local physical conditions.
Self-consistently, 
feedback into the intergalactic medium (IGM)
from young stars in the ``galaxies"
is allowed, in three related forms:
supernova thermal energy output,
UV photon output and mass ejection from the supernova explosions. 
The model reproduces
the observed UV background and metallicity distributions among others.

\section{Results}

Before describing our detailed results,
it is useful
to give a simple and surprisingly accurate order of magnitude
physical argument.
In all standard pictures for the growth of structure there is 
imprinted at an early time a spectrum of perturbations,
with the amplitude of the fluctuations
being largest at small scales and smaller at larger scales.
After decoupling at $z\sim 1000$, all waves
grow (due to self-gravity)
roughly as $(1+z)^{-1}$.
They have
reached the nonlinear length and mass scales 
of $5.6h^{-1}$Mpc and 
$7.5\times 10^{13}h^{-1}\msun$, respectively, by $z=0$
(our fiducial model is adopted for this illustration).
When a perturbation of a given scale $L$ collapses at time
$t$ due to gravity, geometry
indicates it must do so with a velocity
$v\approx L/t$ and, as opposite sides of the collapsing
perturbation meet and attempt 
to cross one another, a shock is generated behind which 
the sound speed 
is $C\approx v$.
Combining these simple arguments and noting that $t^{-1}\approx H$
gives us
\begin{equation}
C_z^2 = K (H L_{nl})_z^2,
\end{equation}
\noindent 
where $K$ is a numerical constant
and $H_z$ and $L_{nl,z}$ are
the Hubble constant and nonlinear length scale at epoch $z$, respectively.
Applying this to the current epoch (with $K=1$)
gives $C_o=565$km/s, which corresponds to a temperature
of $3.9\times 10^7$K.
This should correspond to the typical
temperature of recently collapsed
(i.e. high density) objects,
with the global mean temperature somewhat lower.
Note that $K$ is only a dimensionless number 
to indicate the nature of the scaling
relation and is not intended to model
shock jump conditions.

Figure 1 shows the mean computed 
temperature as a function of redshift from the $[100h^{-1}$Mpc$]^3$
simulation.
The volume weighted
average temperature at $z=0$ is $10^{5.5}$K,
the mass ($\rho$) weighted 
average temperature is $10^{7.4}$K
(the value given by equation 1 with $K=1$)
and the $\rho^2$ weighted 
average (roughly speaking emission weighted) is $10^{8.0}$K.
Thus, the high density regions are 
in just the temperature domain indicated
by equation (4)
with the constant taken to be unity.
Figure 1 shows that equation (4) (adopting $K=1$)
in fact roughly tracks the density weighted
averages from redshift three onwards,
confirming the simple physical picture presented earlier.

Now let us examine the results in greater detail,
dividing the gas into three temperature 
ranges (1) $T > 10^7$K (the normal X-ray emitting gas, predominantly
in collapsed and virialized clusters of galaxies);
(2) $10^7~$K$>T>10^5~$K gas, 
which we will call the warm/hot gas and is mainly in unvirialized 
regions;
(3) $T<10^5~$K warm gas, 
which is seen
in optical studies both in absorption and emission.
A last component (4) is the cold gas that has been condensed
into stellar objects, which we designate ``galaxies",
and which will
contain stars and cold gas.
Figure 2 shows the evolution of these four components,
and the results are consistent with our other knowledge.
Most of the volume is always filled with warm
(Lyman alpha forest) gas, but the mass fraction
in this component declines from $94\%$ at $z=3$ to $26\%$ at $z=0$,
consistent with the HST observed clearing of the forest
and of low-z redshift $\lya$ cloud gas (Shull 1996,1997).
Note that at low reshift denser gas in and around galaxy halos
may make significant contributions to 
the observed low redshift $\lya$ clouds,
which compensates somewhat the decrease of the warm component
and incidentally leads to a leveloff of $\lya$ clouds at low redshift
(Bahcall \etal 1996).
The hot component increases in mass fraction, reaching
$12\%$ by mass at $z=0$,
and is consistent with observations of the local
properties of the X-ray emitting great clusters 
(e.g., Cen \& Ostriker 1994; Bryan \& Norman 1998).
The condensed component remains small 
(12\% and is possibly overestimated
in this simulation), consistent with the known mass density in galaxies.

But we wish to focus attention on the solid circles in Figure 2b:
the warm/hot gas rises dramatically in abundance
with increasing time and dominates
the mass balance by $z=0$,
reaching $52\%$ of the noncondensed mass fraction
or 47\% of the total baryons.

Figure 3 shows the spatial 
distribution of this gas. We see 
a filamentary structure (in green) where,
at the high density nodes (red), 
``galaxies" have formed or been collected.
The spectral features of this warm/hot gas
are in the EUV
and soft X-ray,  
which are very difficult to observe at low redshift
because of the interstellar medium in the Galaxy.
In the soft X-ray the Galaxy
is a strong source of emission at an effective temperature of $\sim 10^6$K,
whereas the neutral hydrogen in the Galaxy
prevent observations of the EUV at $\lambda < 912$A.

Finally, Figure 4 shows the distribution of emission of the gas,
which convolves the density weighted 
distribution with the familiar cooling curve $\Lambda(T)$.
We see, at $z=0$,
two peaks, one at $10^8$K and a higher
and broader one at $10^5$K,
with a broad flat valley in between.
Note that cooling due to metals is included in the calculation,
where metals are produced self-consistently in the simulation.
The computed metallicities are consistent with observations
where comparisons can be made.

\section{Discussion}

How robust are these results to variation of model assumptions?
We believe that any model which is consistent
with other well measured local universe observations
would give essentially the same result.
The reason for this is simply that the temperature is determined
by the nonlinear mass scale (equation 4),
and that, in turn is close to the $8h^{-1}$Mpc
scale which fixes the abundance of rich clusters and
is fairly robustly determined to a narrow range (e.g., Cen 1998):
$\sigma_8 \approx (0.50\pm 0.05)\Omega_0^{-0.40}$
for $0.2<\Omega_0<1.0$.
Only a very small extrapolation
is needed to go from this well observed scale to the nonlinear scale,
so the estimated temperature of collapsed regions
that we find will be common to all models based on the
gravitational growth of structure - 
as normalized to local cluster observations.
In fact, analysis of 
our earlier work (Ostriker \& Cen 1995) covering
nearly a dozen different models
(at lower resolution)
has shown that from one-half to two-thirds
of all baryons in all models
are consistently and robustly in this warm/hot gas at $z=0$.
We also show in Figure 2b the warm/hot component for
two other models, an open CDM model
with $\Omega_0=0.40$ and $\sigma_8=0.75$ (dotted curves), and
a mixed hot and cold dark matter model with 
$\Omega_{hot}=0.30$ and $\sigma_8=0.67$ (dashed curves)
computed completely independently
by Bryan \& Norman (1998).
Yet, their results are in excellent agreement with ours.
The density fluctuation amplitude normalization of 
their mixed dark matter model 
is somewhat below that required to produce the abundance
of local galaxy clusters.
Therefore, an appropriately normalized mixed dark matter model
would yield the warm/hot gas fraction in still better agreement with 
the other two models.

How robust are these results to the input modeling physics?
In addition to gravity and hydrodynamics,
there are three major pieces of relevant input physics:
the meta-galactic radiation field,
energy deposition into IGM from young galaxies 
and metal cooling.
All these three pieces of physics are already included
in the simulations examined here.
Let us discuss them in turn.
First, changing the meta-galactic radiation field would not
make any significant difference  
to the baryons at the relevant temperature and density ranges,
because the warm/hot gas is heated primarily by collisions.
Second, energy feedback from young galaxies
is secondary compared to shock heating due to structure collapse.
An estimate can be made:
the equivalent kinetic energy input per unit mass
due to supernova energy feedback, averaged over all baryons,
is $f_{gal} e_{SN} c^2 \sim  (0.1) \times (1.0 \times 10^{-5}) \times c^2 = 300^2$(km/s)$^2$, or $2\times 10^6$K, 
[where $f_{gal}\sim 0.1$ is the mass fraction of baryons that 
is formed into galaxies by $z=0$ (open squares in Figure 2b),
and $e_{SN}$ is the fraction of energy that
supernovae deposit into IGM in terms of total rest mass],
which is compared to the mass-weighted 
temperature at z=0, $\sim 2\times 10^7$K (dotted curve in Figure 1).
Thus, it seems that supernova feedback contributes about $10\%$
in the overall energy budget.
Finally, the metal cooling time is
$t_{cool} = kT/n\Lambda=4.0\times 10^{-5} T_5/\delta \Lambda$,
where $T_5=T/10^5$, $\delta$ is the density of the gas in units
of the global baryonic mean
and $\Lambda$ is the cooling rate (Bohringer \& Hensler 1989).
Assuming a typical metallicity of $0.1$ solar,
we obtain $t_{cool}=(1.2,54.0,675.0) t_H/\delta$,
at $T=(10^5,10^6,10^7)$K, respectively,
where $t_H$ is the current Hubble time.
For most of the warm/hot gas, $\delta$ is about a few.
Therefore, most of the warm/hot gas has not cooled significantly 
and will not be able
to cool in the next Hubble time, especially considering
the steady input of gravitational energy.
A small fraction of gas
at the lower end of the temperature range ($T<10^5$)
has, of course, cooled to form galaxies 
and a smaller fraction 
will continue to cool to form more galaxies in the next Hubble time.
But most of the gas, including part of the gas that may otherwise
be able to cool, will likely be incorporated into larger and hotter
systems, due to merger as well as breaking of still longer waves.

How can this gas be observed?
Cen \etal (1995) argued, from a less accurate computation
than the present one,
that a significant fraction of the soft X-ray background 
$<1.0~$keV was due to this warm/hot gas
and that this could be verified by association of this soft X-ray
background radiation field
with relatively nearby large scale structure features,
as well as by the soft X-ray angular auto-correlation function
(Soltan \etal 1996).
A more accurate prediction is now underway.
Recent ROSAT observations of the soft X-ray background (Wang \& McCray 1993)
seem to hint the existence of this warm/hot gas.
Another useful method of detecting the soft X-ray background
is to search for shadows cast by nearby neutral hydrogen-rich
galaxies (Wang \& Ye 1996; Barber, Roberts, \& Warwick 1996)
using AXAF.
Second, Hellsten \etal (1998), on the basis of simulations similar 
to those reported on here,
have predicted the existence of an X-ray absorption forest due to
ionized oxygen (O VII 574~eV line)
in the warm/hot temperature range.
Perna \& Loeb (1998) have also made similar calculations based on
simplified models for the IGM.
Work underway with T. Tripp and E. Jenkins indicates
that vacuum UV absorption lines 
(OVI 1032A,1038A doublets)
should be detectable by current instruments
due to gas in this temperature range
primarily in the distant outskirts of galaxies.
Third, strong soft X-ray emission lines from highly ionized species
(such as O VIII 653~eV line)
should also be observable (Jahoda \etal 1998, in preparation).
Fourth, recent work on
EUV observations (Mittaz, Lieu, \& Lockman 1998)
may indicate emission structures coincident
with or projected onto rich X-ray clusters due to 
the warm/hot component as seen Figure 3.
Fifth, the warm/hot gas may show up as very broad, relatively
weak (mostly having $N_{HI}\le 10^{13}$cm$^{-2}$ with a 
small fraction at higher column densities),
low redshift $\lya$ clouds 
but observations more sensitive than current ones (Shull 1996,1997)
are required.
Finally, this gas may be detected through
cross-correlating soft X-ray background (Refregier, Helfand, \& McMahon 1997)
or Sunyaev-Zel'dovich effect (Refregier, Spergel, \& Herbig 1998) with
galaxies.

Overall, on the basis of 
preliminary computations (by scaling $\Omega_{baryon}$ down from 
$0.049$ to $0.037$ using latest observations and 
noting the $J\propto \Omega_{baryon}^2$ scaling relation),
we expect that $1/4$ of the extragalactic soft X-ray background 
(at $0.7~$keV) comes from
this diffuse warm/hot gas: $J_{WH}=7~$keV/sec/cm$^2$/keV/sr
is compared to the Wang-Ye (1996) estimate of 
$\ge 4~$keV/sec/cm$^2$/keV/sr
for the diffuse extragalactic SXRB.
Half is emitted by structures at redshift $z<0.65$ and
three-quarters from $z<1.00$; so one should
be able to identify the optical features associated 
with the emitting gas.
In sum, several experiments using existing or planned 
facilities have the possibility of detecting, in a relatively
precise way, the abundance and cosmic distribution of this warm/hot
gaseous component of the universe,
which is likely to contain most of the baryonic matter
at the current epoch.

\acknowledgments

We thank Richard Mushotzky,
Jim Peebles and David Spergel for useful discussions,
and Greg Bryan for allowing us to show his unpublished results.

\clearpage
\figcaption[FLENAME]{
shows globally averaged temperatures
within our $[100h^{-1}Mpc]^3$
box as a function of redshift from the simulation.
The solid, dotted and short-dashed curves
are the average temperatures weighted
by volume, density and density squared, respectively.
The long-dashed curve represents the results from
the simple physical argument as indicated by equation 4 with
the constant $K=1$,
and indicates that the temperature of the high density nonlinear
regions is well represented by that formula.
\label{fig1}}

\figcaption[FLENAME]{
shows the evolution of the four components of cosmic baryons
(see text for definitions).
Panel (a) shows the volume fractions of the four components
and 
panel (b) shows the mass fractions.
Examination of (2b) shows that more than half
of the baryons at redshift zero are in the temperature range
$10^7$K$>T>10^5$K.
Also shown are the warm/hot component for
two other models: an open CDM model
with $\Omega_0=0.40$ and $\sigma_8=0.75$ (dotted curves), and
a mixed hot and cold dark matter model with 
$\Omega_{hot}=0.30$ and $\sigma_8=0.67$ (dashed curves).
These two models were computed completely independently
by Bryan \& Norman (1998).
\label{fig2}}

\figcaption[FLENAME]{
shows the spatial distribution of the 
warm/hot gas with temperature in the range 
$10^{5-7}$K at $z=0$.
The green regions have densities about 
ten times the mean baryon density of the universe at $z=0$;
the yellow regions have densities about one hundred
times the mean baryon density, while
the small isolated regions with red and saturated dark colors 
have even higher densities reaching about 
a thousand times the mean baryon density,
and are sites for current galaxy formation.
\label{fig3}}

\figcaption[FLENAME]{
shows the distribution of emission of the gas
which convolves the density weighted 
distribution with the familiar cooling curve $\Lambda(T)$,
including emission from metals where are computed self-consistently
in the simulation.
Three epochs, $z=(0,1,3)$, are shown 
as solid, dotted and dashed curves, respectively. 
At redshift zero most of the emission is from 
the warm/hot gas with
a secondary maximum at the normal hard X-ray peak of $10^8$K.
\label{fig4}}
 
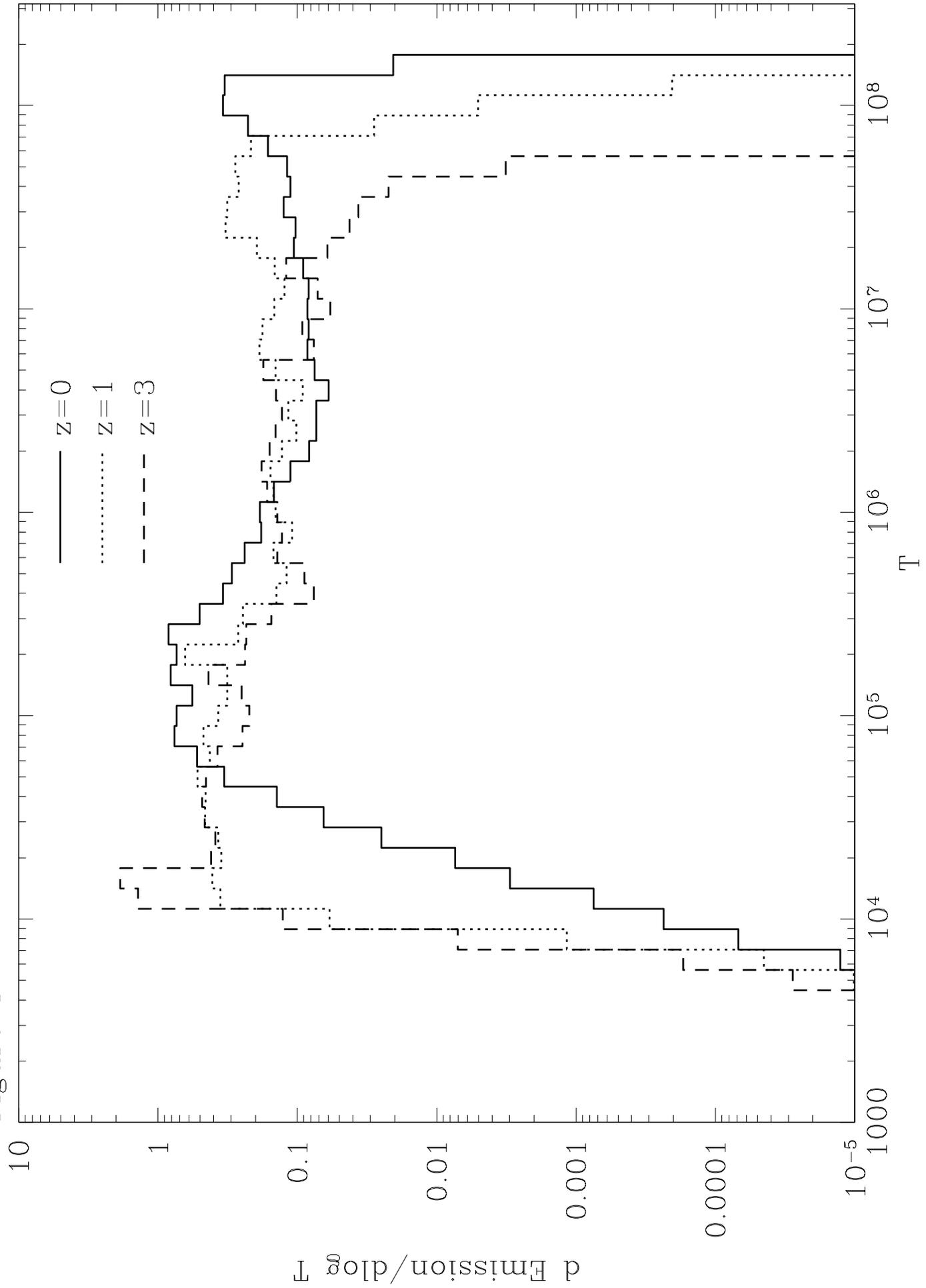
\end{document}